\documentclass[a4paper,fleqn, review, 12pt]{cas-sc}

\usepackage[numbers]{natbib}
\usepackage{bm}
\usepackage{float}
\usepackage{booktabs}
\usepackage{multirow}
\usepackage{adjustbox}
\usepackage{threeparttable}
\usepackage{xcolor}
\usepackage[section]{placeins}
\usepackage{url}
\usepackage{setspace}
\usepackage{soul}
\usepackage{amsmath}

\def\tsc#1{\csdef{#1}{\textsc{\lowercase{#1}}\xspace}}
\tsc{WGM}
\tsc{QE}
\tsc{EP}
\tsc{PMS}
\tsc{BEC}
\tsc{DE}

\begin{document}
\let\WriteBookmarks\relax
\def\floatpagepagefraction{1}
\def\textpagefraction{.001}

\shorttitle{N-of-1 Trials for Gait}

\shortauthors{L. Zhou et~al.}

\title [mode = title]{Analyzing Population-Level Trials as {N}-of-1 Trials: an Application to Gait}                      



%
\author[1]{Lin Zhou}[type=editor,
                        orcid=0000-0001-9916-3878]

\cormark[1]

\author[2]{Juliana Schneider}[orcid=0000-0002-1447-6515]

\author[1]{Bert Arnrich}[orcid=0000-0001-8380-7667]

\author%
[2,3,4]{Stefan Konigorski}[orcid=0000-0002-9966-6819]
\cormark[1]


\affiliation[1]{organization={Digital Health - Connected Healthcare, Hasso Plattner Institute, University of Potsdam},
    city={Potsdam},
    country={Germany}}
    
\affiliation[2]{organization={Digital Health \& Machine Learning, Hasso Plattner Institute, University of Potsdam},
    city={Potsdam},
    country={Germany}}

\affiliation[3]{organization={Department of Statistics, Harvard University},
    city={Cambridge},
    country={USA}}

\affiliation[4]{organization={Hasso Plattner Institute for Digital Health at Mount Sinai, Icahn School of Medicine at Mount Sinai},
    city={New York},
    country={USA}}

\cortext[cor1]{Corresponding author}



\begin{abstract}
Studying individual causal effects of health interventions is important whenever intervention effects are heterogeneous between study participants. Conducting N-of-1 trials, which are single-person randomized controlled trials, is the gold standard for their analysis. As an alternative method, we propose to re-analyze existing population-level studies as N-of-1 trials, and use gait as a use case for illustration. Gait data were collected from 16 young and healthy participants under fatigued and non-fatigued, as well as under single-task (only walking) and dual-task (walking while performing a cognitive task) conditions. As a reference to the N-of-1 trials approach, we first computed standard population-level ANOVA models to evaluate differences in gait parameters (stride length and stride time) across conditions. Then, we estimated the effect of the interventions on gait parameters on the individual level through Bayesian repeated-measures models, viewing each participant as their own trial, and compared the results. The results illustrated that while few overall population-level effects were visible, individual-level analyses revealed differences between participants. Baseline values of the gait parameters varied largely among all participants, and the effects of fatigue and cognitive task were also heterogeneous, with some individuals showing effects in opposite directions. These differences between population-level and individual-level analyses were more pronounced for the fatigue intervention compared to the cognitive task intervention. Following our empirical analysis, we discuss re-analyzing population studies through the lens of N-of-1 trials more generally and highlight important considerations and requirements. Our work encourages future studies to investigate individual effects using population-level data.
\end{abstract}




\maketitle

\section{Introduction}
\label{sec:introduction}
In the majority of research studies, the focus lies on identifying average effects in a population of individuals, such as in large cohort studies or randomized controlled trials (RCTs). However, especially if there are heterogeneous individual effects, it can be of great interest to investigate associations on an individual level. Estimating and testing these individual effects pose challenges. One approach is to employ statistical or machine learning models to estimate individual effects from the population-level studies. To this end, different methods have been proposed in recent years~\cite{bica2021real, alaa2017bayesian, shalit2017estimating, Lee2018}. As another approach, a new study can be designed with the specific aim of investigating individual-level effects. For this, the study design of so-called N-of-1 trials has been established as the gold standard~\cite{nikles_essential_2015}. In N-of-1 trials, the effect of one or more interventions is investigated in an individual by measuring the outcome of interest over time across alternating phases in which the interventions are applied. As such, N-of-1 trials are multi-crossover single-person RCTs ~\cite{lillie2011n, mirza2017history}. As a third approach, which we propose in this study, population-level data can be re-analyzed through the lens of N-of-1 trials.
This approach can be generally applied to experimental studies that use a repeated-measures design in which two or more conditions with interventions are measured repeatedly. We illustrate this for one specific study in the following. 

In order to analyze individual data as N-of-1 trials, it is important that multiple measurements are available for each individual. This can be aided if sensors are used that capture data in short intervals. For example, heart rate variability can be calculated within time windows of several minutes, and continuous recording on a smartwatch can generate a large number of measurements for analysis~\cite{ishaque2021trends}. Another example is gait analysis, which is the study of walking patterns. Devices such as instrumented walkways, or wearable sensors can measure hundreds of strides in minutes~\cite{trautmann2021tripod}. There are various other use cases where our proposed analysis method may apply, for illustration, here we focus on an application to gait. 

Gait can be quantified using spatiotemporal parameters, such as stride length, stride time, speed, or cadence. These gait parameters provide crucial insights into a person's health status. For example, gait speed and variability have been associated with life expectancy and risk for falls in older adults~\cite{stanaway2011fast, HAUSDORFF20011050}. Although gait is typically regarded as an isolated and highly automatic task, evidence suggests that gait patterns differ when concurrently performing a secondary task (e.g., cognitive or motor interference task). Such dual-task situations, which closely mimic daily life walking~\cite{hillel2019every, BAYOT2018}, have been associated with slower gait speeds and increased stride times ~\cite{Ebersbach1995, Nohelova2021, SMITH2016, Montero-Odasso2012}. Consequently, studying gait under these conditions may provide clinically relevant insights into gait modulations in daily life. 

A more comprehensive understanding of gait in real-life walking should also consider the aspect of physical fatigue because it knowingly affects gait kinematics and kinetics, and is linked to a higher risk of slip-induced falls as well as impaired movement controls in young and old healthy adults~\cite{parijat2008effects, santos2019effects, hatton2013effect}.  Existing studies investigating the effects of muscle fatigue on gait performance have reported heterogeneous outcomes. For example, for young healthy adults, Granacher~et~al.~\cite{granacher2010effects} observed statistically significant decreases in gait speed and stride length in a fatigue condition, while  Parijat~et~al.~\cite{parijat2008effects} reported no statistically significant changes in gait speed. In older healthy adults, muscle fatigue only resulted in rather moderate changes in gait parameters~\cite{santos2019effects}. Regarding stride length, some studies reported an increase~\cite{granacher2010effects, Barbieri2014, morrison2016walking}, while others reported no changes~\cite{hatton2013effect, TOEBES2014, Helbostad2007}.

One possible explanation for the aforementioned discrepancy in the observed effects of fatigue could be that the group-level analyses typically performed in gait studies do not capture the heterogeneous gait changes among individuals. The above-mentioned studies did not investigate heterogeneity among their participants, and none of the studies included covariates such as gender, height or weight in their analysis. However, it is known that gait characteristics are highly individualized and persist for a long period. As reported by Horst~et~al., classification accuracy for identifying 46 healthy individuals using their gait patterns remained at 99\% for one year ~\cite{HORST2017}. Moreover, there is evidence that gait changes in response to interventions, such as athletic training or disease treatment, are also individualized~\cite{chan2020effects, marxreiter2018sensor, nonnekes2018towards}. 

The highly individualized nature of gait and gait modifications suggests that individual-level analyses could provide insights that are not evident from population-level analyses. In the context of the effects of fatigue and dual-task on gait, studying individual gait responses can allow assessing and monitoring an individual’s risk for falls and mobility impairments. Typically, statistical models have accounted for inter-personal differences by including covariates, or by including random effects to better approximate each individual’s own average gait response. In these cases, the estimand is still a population average and the individual responses are not characterized. Therefore, N-of-1 approaches are needed to adequately characterize the individual response. However, to the best of our knowledge, only one series of N-of-1 trials has been conducted on gait, in which Maguire~et~al. compared the effect of different walking aids on gait and balance for chronic stroke patients and revealed different responses across the participants~\cite{maguire2020replacing}. 


Here, we investigate how existing data from population-level studies can be re-analyzed through the lens of N-of-1 trials to estimate individual-level effects. To this aim, we use data from a population-based study in a repeated-measures design that investigated the effects of physical fatigue and cognitive task on gait~\cite{zhou2023duo}. We estimate personalized gait parameters (stride length and stride time) from Bayesian repeated-measures models, and compare the results with a standard population-level ANOVA model. Finally, we discuss re-analyzing population studies through the lens of N-of-1 trials more generally and highlight important considerations and requirements.

\section{Materials and methods}
\subsection{Overview of the gait study}
Sixteen young healthy participants (eight males, eight females) were enrolled in the study. Eligibility for the study was determined using the Physical Activity Readiness Questionnaire (PAR-Q) and only participants without medical restrictions for performing physical activities (i.e. with all negative responses) were allowed to take part in the study. At the first visit (see study design below), personal characteristics were assessed, including the physical activity levels of the participants on a scale of 1~(low)~-~3~(high) using the International Physical Activity Questionnaire (IPAQ)~\cite{craig2003international}.

\begin{figure}[ht]
    \centering
    \includegraphics[width=0.8\textwidth]{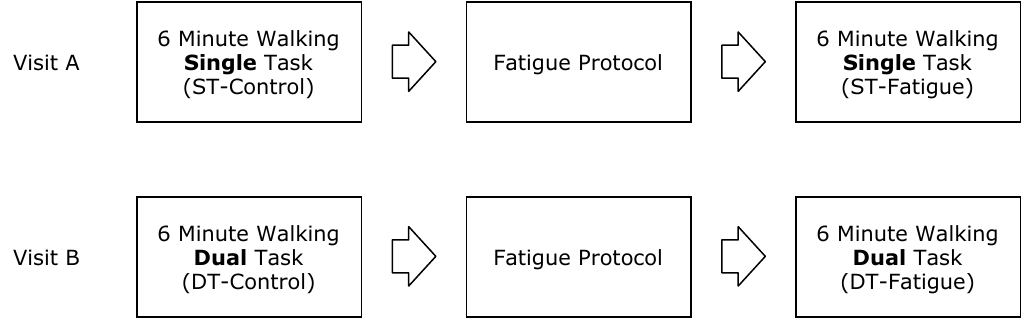}
    \caption{Study design. The study consisted of two visits which were seven days apart. The order of visits A and B was randomized among participants. During each visit, the participants performed two 6-minute walking assessments, separated by a fatigue protocol. During visit B, the participants performed a number-counting task while walking. ST: Single-Task (only walking), DT: Dual-Task (walking while counting numbers).}
    \label{fig:exp_flow}
\end{figure}

Figure~\ref{fig:exp_flow} shows an overview of the study design. The study consisted of two visits, referred to as visits A and B in the following, which were seven days apart. The order of A and B was randomized among participants, and at each of the two visits, the participants performed two walking assessments, separated by a fatigue protocol. During visit A, participants first completed a 6-minute walking assessment in a corridor with a distance of 35 meters in one direction. Then, the participants performed a repeated sit-to-stand all-out fatigue protocol to induce muscular fatigue in the lower limbs. Immediately following the fatigue protocol, participants repeated another 6-minute walking assessment. During visit B, the experimental procedure was the same as in visit A, except that the participants had to count numbers while walking (i.e., the dual-task condition). Details of the fatigue protocol and the cognitive task are described in Supplementary~Text~A.

Our primary focus of the measurements obtained in the walking sessions was on two gait parameters: stride length and stride time. These parameters were captured using inertial measurement units (IMUs) attached to the participants' shoes, which measured tri-axial acceleration and angular velocity of the foot movement. Details of the IMU gait analysis methods are described in Supplementary~Text~A. In total, gait parameters from four walking sessions were collected for each participant: single-task control (ST-Control), single-task fatigue (ST-Fatigue), dual-task control (DT-Control) and dual-task fatigue (DT-Fatigue).  Hence, each participant had either the intervention sequence ST-Control – ST-Fatigue – DT-Control – DT-Fatigue, or the sequence DT- Control – DT-Fatigue – ST-Control – ST-Fatigue. Hundreds of gait measurements were made during each block.

The study was approved by the ethics committee of the University of Potsdam (number 63/2020), and all experiments were conducted according to the latest revision of the Declaration of Helsinki. All participants provided written consent prior to data collection.

\subsection{Statistical analyses}
\subsubsection*{Descriptive statistics and population-level analyses}
As a first step, we computed descriptive statistics of age, body weight, height, and the IPAQ physical activity level of all participants. Next, we performed a population-level analysis using a two-way repeated measures ANOVA to serve as a reference for the comparison to the N-of-1 trial analyses. In ANOVA, we used stride length and stride time as outcomes and tested for the effect of physical fatigue and cognitive tasks, which were included as fixed factors. In addition, age, body height and body mass were included as covariates in the model. The model is described in detail in Supplementary~Text~B.

\subsubsection*{N-of-1 trials analysis using Bayesian repeated-measures models}
\label{sec:bayesian_model}
In our main analysis, we analyzed the data through the lens of N-of-1 trials. For each participant, we estimated the individual marginal effects of the physical fatigue intervention and cognitive intervention on the gait parameters stride length and stride time.
In contrast to typical N-of-1 trials with multiple crossovers, the data from our study consists of four blocks of repeated measurements of the outcome gait parameters for each participant. This 2~$\times$~2 experimental design is typical in population-level studies~\cite{granacher2010effects}. 

We used Bayesian linear repeated-measures models to fit probabilistic models of the data distribution to the gait time series data, separately for each participant, and separately for stride length and stride time. Such Bayesian models provide a probabilistic description of the data for interpretation~\cite{makowski2019indices} and allow for the incorporation of prior knowledge, which is not as easily or transparently done in conventional frequentist analysis. A model with a first-order autoregressive (AR1) error structure was used, which acknowledges that (for the same person) the covariance between errors from the observations may not be equal, and decreases towards zero with increasing lag. The AR1 error structure has been recommended in the literature for single-subject time series data in the context of Bayesian hypothesis testing~\cite{deVries2013bayesian}. 

For a more detailed description, let $y_i, i = 1, \dots n$, denote the $i$-th observation (i.e., stride time or stride length of the $i$-th stride) of a participant in the study. It is worth noting that the total number of observations $n$ varies for each participant. For ease of notation, we use $n$ in the following in our description of the linear model that is fit for each individual separately:
\vspace{-20pt}
\begin{equation}
    \label{eq:linear_model}
    y_i~=~ \beta_1 + \beta_2 X_{i2} + \beta_3 X_{i3} + \beta_4 X_{i2}\cdot X_{i3} + \epsilon_i = \bm{X_i}\bm{\beta} + \epsilon_i, 
\end{equation}
\vspace{-20pt}

\noindent where $\bm{X}$~=~($X_{1}$, $X_{2}$, $X_{3}$, $X_{2}\cdot X_{3}$) is the~($n~\times~4$) design matrix. $X_1 = \bm{1_n}$ is a vector of ones of length~$n$ that represents the intercept. $X_2$ denotes whether the individual was in the cognitive task condition (dual task, $X_{i2}=1$) or not (single task, $X_{i2}=0$), and $X_3$ denotes whether the individual was in the fatigue condition ($X_{i3}=1$) or in the control condition ($X_{i3}=0$). Further, $X_{2}\cdot X_{3}$ denotes the interaction between cognitive task and fatigue. $\bm{\beta}~=~(\beta_1, \beta_2, \beta_3, \beta_4)^T$ is a vector of fixed effects. 
Hence, the mean gait parameters under the four walking conditions were estimated using the following combinations of $\bm{\beta}$ coefficients. ST-Control: $\beta_1$, DT-Control: $\beta_1$ + $\beta_2$, ST-Fatigue: $\beta_1$ + $\beta_3$, DT-Fatigue: $\beta_1$ + $\beta_2$ + $\beta_3$ + $\beta_4$.
$\bm{\epsilon}$ represents the error drawn from a multivariate normal distribution: 

\vspace{-20pt}
\begin{equation}
   \label{eq:multivariate_normal_distribution}
   \bm{\epsilon} \sim MVN(\bm{0}, \sigma^2 \bm{\Psi})
\end{equation}
\vspace{-20pt}

\noindent where $\sigma^2$ is the error variance, and \bm{$\Psi$} is a variance-covariance matrix determined by the AR1 process~\cite{deVries2013bayesian}, as in Equation~\ref{eq:AR1_Sigma}: 

\vspace{-20pt}
\begin{equation}
    \label{eq:AR1_Sigma}
    \bm{\Psi}~=~\frac{1}{1-\phi^2}  
    \begin{bmatrix}
        1               & \phi        &  \phi^{2}   & ...    & \phi^{n-1}    \\
        \phi            & 1           &  \phi       & ...    & \phi^{n-2}    \\
        \phi^{2}        & \phi        &  1          & ...    & \phi^{n-3}    \\
        \vdots          & \vdots      & \vdots      & \ddots & \vdots        \\
        \phi^{n-1}      & \phi^{n-2}  & \phi^{n-3}  & ...    & 1             \\
    \end{bmatrix}
\end{equation}
\vspace{-20pt}

\noindent Here, $\phi$ is the correlation coefficient between two successive time points. The exponent of $\phi$ reflects the decline of the correlation between time points according to their distance from one another.

The Markov Chain Monte Carlo (MCMC) method with Gibbs sampling was used. 
Among all parameters of the model, the parameter of primary interest is the vector $\bm{\beta}$, since combinations of its elements make up the mean gait parameter distributions for the four walking conditions (i.e., ST-Control, ST-Fatigue, DT-Control and DT-Fatigue). 
While informative prior distribution for the parameter $\beta_1$ can be directly inferred from studies on normal gait parameters of young healthy adults~\cite{bernal2016reliability}, there is not enough information available to assume priors for the other parameters. As a result, we chose to use non-informative priors in the main analyses. We assumed a half-Cauchy distribution for the residual variance $\sigma$ (cf. Equation~\ref{eq:multivariate_normal_distribution})
in the Bayesian model as described in~\cite{gelman2006prior}, with default priors recommended by Gelman~et~al.~\cite{gelman2008weakly}. In sensitivity analyses, were tested different informative priors, the details are described in the "Sensitivity analyses" section below. Table~\ref{tab:priors} provides an overview of the priors. 

In the sampling procedure, we used 2 chains, 5000 burn-in steps, 1 thinning step (i.e. no thinning), and 10,000 iterations. To reduce the amount of computation, data used for the AR1 model were taken only from the left foot, and down-sampled by a factor of five (i.e., selecting every fifth sample sequentially). We confirmed with visual inspection that down-sampling did not change the overall distribution of the data. In addition, there was no obvious change in the gait parameters over time with each walking condition for the same person. Example plots of the data can be found in Supplementary~Text~A.

From the Bayesian models and conditional parameters, our aim was to estimate the marginal effects of physical fatigue or cognitive task, which we expect to vary among the participants. To estimate the probability of the marginal effect being clinically meaningful, we compared the effect to a pre-defined threshold~\cite{stunnenberg2018effect}. More concretely, the marginal effects of fatigue or cognitive task were derived as a function of the $\bm{\beta}$ coefficients of the AR1 model (see details in Supplementary~Text~B). Then, the posterior probability distribution of the marginal effects was estimated along with other model parameters in the AR1 model. Subsequently, we defined the probability of the marginal effect being meaningful as the posterior probability of the estimated marginal effect being greater than a given threshold (i.e., the area under the curve of the posterior distribution bounded below by this threshold, following Stunnenberg et al.~\cite{stunnenberg2018effect}). Throughout the text, we refer to this as the posterior probability of a meaningful effect or change (e.g., as in Figure~\ref{fig:posterior_prob_meaningful_change}).
For our example dataset, the thresholds for meaningful change were determined based on previous studies on the effects of fatigue in a similar population. We defined clinically meaningful effects on stride length and stride time to be~3\% and~2\%, respectively~\cite{granacher2010effects, Barbieri2014}. This can be generalized to a variety of settings to investigate the probability of meaningful effects for an individual. Future studies adopting this method should define the meaningful effect threshold based on the specific domain context.

The convergence of the MCMC chain was confirmed with potential scale reduction factor (PSRF) and trace plots. MCMC chain resolution was evaluated using the effective sample size (ESS), which measures the efficiency of Monte Carlo methods such as MCMC~\cite{martino2017effective}. More details on the MCMC diagnostics and on their results can be found in Supplementary~Text~C.
To confirm that the posterior estimates accurately represent the observed data, a posterior predictive check was performed by comparing the posterior distributions with the distribution of the observed samples. More specifically, the posterior distribution of the intercept and effects were used to reconstruct the modeled distributions of gait parameters under the four conditions. These modeled distributions are then compared with the observed sample distributions using boxplots. More details are described in Supplementary Text~C.
The Bayesian analysis was performed using JAGS version 4.3.0, run from R version 4.1.1 (R Project for Statistical Computing). Formal specifications of the JAGS models can be found in Supplementary~Text~B. The data and R scripts used for running the analysis can be found at \url{https://github.com/HIAlab/gait_nof1trials}.

\subsubsection*{Sensitivity analyses} 
\label{subsection_sensitivity}
To test how well alternative Bayesian models can estimate the posterior distribution, we implemented two additional models, a simple basic model and a time covariate model. In contrast to the AR1 model introduced in Section~\ref{sec:bayesian_model}, both these models assumed that the errors are independent and identically distributed.
As a basic model, we implemented a simple Bayesian fixed effects model with the two fixed factors of fatigue exposure and cognitive task type. Similar to the AR1 model, we assumed a linear relationship and normally-distributed errors, but we assumed here 
that each data point, namely, each stride from the same recording session, is independent of each other. The model structure is identical to that described in Equation~\ref{eq:linear_model}, except that the error term follows a normal distribution with a diagonal covariance matrix $\bm{I}$: $\bm{\epsilon}~\sim~N(\bm{0}, \sigma^2\bm{I})$.

As a second alternative model based on the basic model, we included a linear time trend 
by appending an incremental integer array to the design matrix:

\vspace{-20pt}
\begin{equation}
    y_i~=~ \beta_1 + \beta_2 X_{i2} + \beta_3 X_{i3} + \beta_4 X_{i2}\cdot X_{i3} + \beta_5 C_i + \epsilon_i = \bm{X'_i}\bm{\beta} + \epsilon_i,
\end{equation}
\vspace{-20pt}

\noindent where $C~=~(1, 2, ..., n_1, 1, 2, ..., n_2, 1, 2, ..., n_3, 1, 2, ..., n_4)^{T}$ is a vector of incremental integers with length $n$. The integers re-start from one for each of the four walking conditions. Concretely, $n$ = $n_1$ + $n_2$ + $n_3$ + $n_4$, where $n_1$, $n_2$, $n_3$, $n_4$ are the number of samples under the four walking conditions, respectively. Apart from the linear time trend covariate, the model structure was identical to the basic model. It is worth noting that we estimate one coefficient, namely, $\beta_5$, for the added time trend column $C$. This approach assumes the outcome has the same time trend regardless of walking condition. However, the outcome time trend may in fact differ across the four walking conditions, and this coefficient represents the mean of the time trend under all of these conditions. Future models can accommodate this assumption by including interaction terms between C and each walking condition.

In further sensitivity checks, we compared models based on non-informative and informative priors for all three above-mentioned models. The investigated priors are summarized in Table~\ref{tab:priors}. The distribution of informative priors was based on the corresponding gait parameter values reported for young healthy adults which included a mean stride length of 1.36~m with a standard deviation of 0.08~m, and mean stride time (estimated as doubled step time) of 1.05~s with a standard deviation of 0.06~s~\cite{bernal2016reliability}.

\begin{table}[h]
\caption{Non-informative and informative prior distributions of the Bayesian models.}
\label{tab:priors}
\begin{threeparttable}
\footnotesize
\begin{tabular}{@{}llll@{}}
\toprule
\textbf{Model}                                                     & \textbf{Non-informative Priors}                                                               & \textbf{Informative Priors (SL)}                                                                                                                             & \textbf{Informative Priors (ST)}                                                                                                                             \\ \midrule
\begin{tabular}[c]{@{}l@{}}Basic \& \\ Time Covariate\end{tabular} & \begin{tabular}[c]{@{}l@{}}$\beta_{i \in \{1, 2, 3, 4\}}~\sim~N(0, 10^{-3})$\\ $\sigma~\sim~U(0, 100)$\end{tabular} & \begin{tabular}[c]{@{}l@{}}$\beta_1~\sim~N(1.36, 0.08)$\\ $\beta_{i \in \{2, 3, 4\}}~\sim~N(0, 10^{-3})$\\ $\sigma~\sim~U(0, 100)$\end{tabular}                               & \begin{tabular}[c]{@{}l@{}}$\beta_1~\sim~N(1.05, 0.06)$\\ $\beta_{i \in \{2, 3, 4\}}~\sim~N(0, 10^{-3})$\\ $\sigma~\sim~U(0, 100)$\end{tabular}                               \\ \midrule
AR1                                                                & \begin{tabular}[c]{@{}l@{}}$\beta_{i \in \{1, 2, 3, 4\}}~\sim~N(0, 10^{-3})$\\ $\sigma~\sim~half$-$Cauchy(2.5)$\\ $\phi~\sim~U(-1, 1)$\end{tabular} & \begin{tabular}[c]{@{}l@{}}$\beta_1~\sim~N(1.36, 0.08)$\\ $\beta_{i \in \{2, 3, 4\}}~\sim~N(0, 10^{-3})$\\ $\sigma~\sim~half$-$Cauchy(2.5)$\\ $\phi~\sim~U(-1, 1)$\end{tabular} & \begin{tabular}[c]{@{}l@{}}$\beta_1~\sim~N(1.05, 0.06)$\\ $\beta_{i \in \{2, 3, 4\}}~\sim~N(0, 10^{-3})$\\ $\sigma~\sim~half$-$Cauchy(2.5)$\\ $\phi~\sim~U(-1, 1)$\end{tabular} \\ \bottomrule
\end{tabular}
\begin{tablenotes}
\item SL: stride length, ST: stride time 
\item The default priors are recommended by Gelman~et~al.~\cite{gelman2008weakly}, and the informative priors for stride lengths and stride time are based on the corresponding gait parameter values reported for young healthy adults~\cite{bernal2016reliability}.
\end{tablenotes}
\end{threeparttable}
\end{table}

\section{Results}
\subsection{Characteristics of study participants}
In total, data from sixteen participants (eight males, eight females) were collected for the four walking conditions (ST-Control, ST-Fatigue, DT-Control, and DT-Fatigue). 
The dataset consisted of 3117 strides pooled across all participants. Stride length and stride time from each stride were used as outcome variables in the analyses. The observations were balanced across the all walking conditions and participants, with 788 strides from ST-Control~(49.3~\textpm~3.7 strides per person), 792 strides from ST-Fatigue~(49.5~\textpm~3.7 strides per person), 766 strides from DT-Control~(47.9~\textpm~3.2 strides per person) and 771 strides from DT-Fatigue~(48.2~\textpm~3.5 strides per person). Table~\ref{tab:participant_characteristics} summarizes the participant characteristics, and Table~\ref{tab:stats_ANOVA} summarizes the gait parameters.

\begin{table}[h]
\caption{Participant characteristics.}
\begin{center}
\begin{threeparttable}
\footnotesize
\begin{tabular}{l r r r}
\hline
    \textbf{Variable} & \textbf{Mean~\textpm~SD} & \textbf{Min} & \textbf{Max} \\ 
\hline
Age                         & 27.1 \textpm~3.8    & 21        & 35   \\
Body Mass (kg)              & 71.2 \textpm~12.2   & 54        & 103  \\
Height (cm)                 & 173.8 \textpm~8.6   & 158     & 190  \\
Activity Level\tnote{*}     & 2    & 1 & 3   
      \\ 
\hline
\end{tabular}
\begin{tablenotes}
\item[*] 1, 2, 3 means low, medium, and high activity levels in IPAQ, respectively. The median is reported instead of Mean~\textpm~SD since data contain ordinal values.
\end{tablenotes}
\end{threeparttable}
\end{center}
\label{tab:participant_characteristics}
\end{table}

\subsection{Population-level analysis}
Next, we performed baseline analyses to investigate the population-level effects of physical fatigue and cognitive task on gait. Two-way repeated measures ANOVA indicated very small effects induced by physical fatigue. The main effects of physical fatigue on stride length and stride time had a generalized eta-squared effect size of 0.01 or less (stride length: F(1,15)~=~5.86, p~=~0.03, $\eta^{2}$~=~0.01; stride time: F(1,15)~=~2.56, p~=~0.13, $\eta^{2}$~=~$8.5\times10^{-3}$). The main effects of cognitive task were moderate, with generalized eta-squared effect sizes 0.15 and 0.20 for stride length and stride time, respectively (stride length: F(1,15)~=~18.46, p~=~$6.4\times10^{-4}$, $\eta^{2}$~=~0.16; stride time: F(1,15)~=~21.14, p~=~$3.5\times10^{-4}$, $\eta^{2}$~=~0.22). No statistically significant interaction effects were found (p~=~0.77 for stride length, and p~=~0.99 for stride time). Table~\ref{tab:stats_ANOVA} summarizes the ANOVA results.

\begin{table}[h]
\caption{Descriptive statistics and ANOVA results of the gait parameters from all participants.}
\begin{adjustbox}{max width=\textwidth}
\begin{threeparttable}
\begin{tabular}{@{}lll@{}}
\toprule
 & \textbf{Stride Length Avg (m)} & \textbf{Stride Time Avg (s)} \\ \midrule
ST-Control (n=788) & $1.43 \pm 0.12$ & $1.10 \pm 0.07$ \\
ST-Fatigue (n=792) & $1.41 \pm 0.13$ & $1.09 \pm 0.06$ \\
DT-Control (n=766) & $1.34 \pm 0.11$ & $1.16 \pm 0.07$ \\
DT-Fatigue (n=771) & $1.32 \pm 0.12$ & $1.15 \pm 0.06$ \\
Main Effect Control-Fatigue & F(1,15)~=~5.86, p~=~0.03, $\eta^{2}$~=~0.01 & F(1,15)~=~2.56, p~=~0.13, $\eta^{2}$~=~$8.5\times10^{-3}$ \\
Main Effect ST-DT & F(1,15)~=~18.46, p~=~$6.4\times10^{-4}$, $\eta^{2}$~=~0.16 & F(1,15)~=~21.14, p~=~$3.5\times10^{-4}$, $\eta^{2}$~=~0.22 \\
\bottomrule
\end{tabular}
\begin{tablenotes}
\item ST~=~Single Task, DT~=~Dual Task, Ctrl~=~Control, Avg~=~average. n: total number of strides. F~=~F-value of ANOVA, p~=~p-value of ANOVA, $\eta^{2}$~=~generalized eta-squared (effect size). Summary of gait parameters are expressed as mean~\textpm~standard~deviation.
\end{tablenotes}
\end{threeparttable}
\end{adjustbox}
\label{tab:stats_ANOVA}
\end{table}

\subsection{N-of-1 trials using Bayesian repeated-measures models}
\begin{figure}[h]
     \centering
     \includegraphics[width=0.9\textwidth]{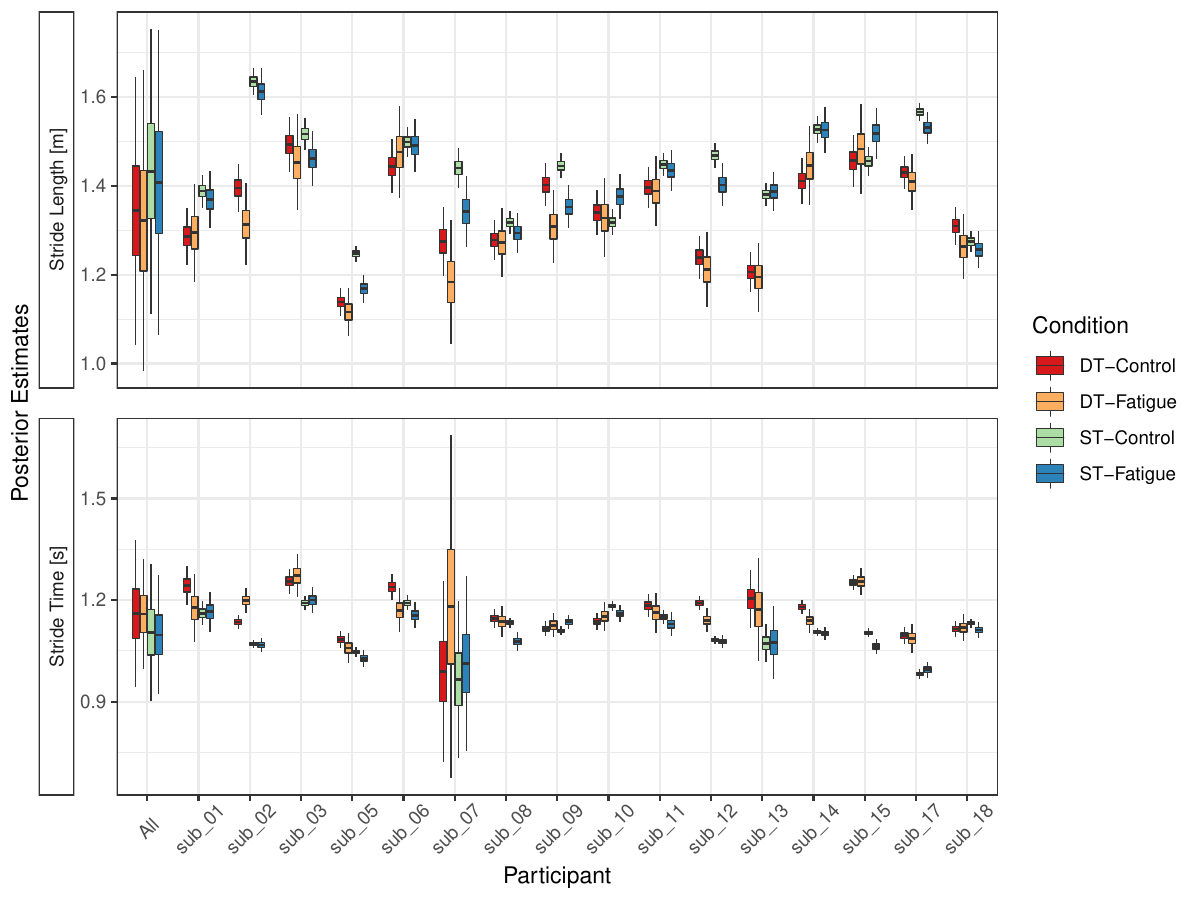}
     \label{fig:boxplot_posterior_SL_ST}
     \caption{Posterior estimates from the AR1 model for stride length and stride time, heterogeneous gait changes could be observed among the participants. Top: Posteriors for stride length; bottom: posteriors for stride times. Distribution of the gait parameters under the four conditions are derived from combinations of the coefficients in the $\bm{\beta}$ vector in the linear model.}
     \label{fig:boxplot_posterior_estimations}
\end{figure}

The posterior distributions for stride length and stride time are illustrated in Figure~\ref{fig:boxplot_posterior_estimations}. A complete summary of the posterior distributions of parameters can be found at~\url{https://github.com/HIAlab/gait_nof1trials/wiki/Data}.
Distributions of the gait parameters under the four conditions are derived from combinations of the elements in the $\bm{\beta}$ vector in the linear model, as described in Section~\ref{sec:bayesian_model}. The results showed that the baseline values of the gait parameters (under the ST-Control condition) varied largely among all participants, and the gait changes under the four walking conditions were also highly heterogeneous among all participants. Figure~\ref{fig:boxplot_posterior_estimations} also shows the aggregated posterior distributions from all participants for reference, which further indicates that the aggregated population-level summary was not a good representation of the highly heterogeneous individual gait effects. 

For stride length, there was a consistent trend among participants that the values from DT conditions were smaller than those from ST conditions, as seen in the population-level ANOVA main effect of the cognitive task described above. Nevertheless, inter-person variation could be observed, for example, as illustrated in Figure~\ref{fig:boxplot_posterior_estimations}, study participant 10 exhibited almost no response under the DT condition compared to the ST condition (mean values changed from 1.34 under ST to 1.33 under DT), whereas participant 2 largely reduced the stride length (mean values changed from 1.62 under ST to 1.35 under DT). 

In contrast, the effects of physical fatigue on stride lengths were smaller on average but more complex on the individual level compared to those induced by the cognitive task, and opposite effects could be observed for different individuals. Especially under DT condition, stride length 
increased from non-fatigue to fatigue condition for participant 6 (from 1.44 to 1.48), participant 14 (from 1.41 to 1.45), and participant 15 (from 1.45 to 1.48) but remained unchanged or decreased for the other participants.

Moreover, the effects of fatigue were larger under DT condition compared to under ST condition for all participants. Similar trends could be observed for stride time, where the DT condition generally induced an increase for all participants, but the individual posterior distributions were heterogeneous. The cognitive task appeared to have increased the variance as well as the effects of fatigue for many participants. It is worth noting that the posterior estimates for participants  7 and 13 had unusually large variations compared to those for all other participants. Quality control analyses revealed that the MCMC chains did not converge for these two participants, more details are presented in section MCMC Chain Convergence in Supplementary~Text~C.

\begin{figure}
    \centering
    \includegraphics[width=0.9\textwidth]{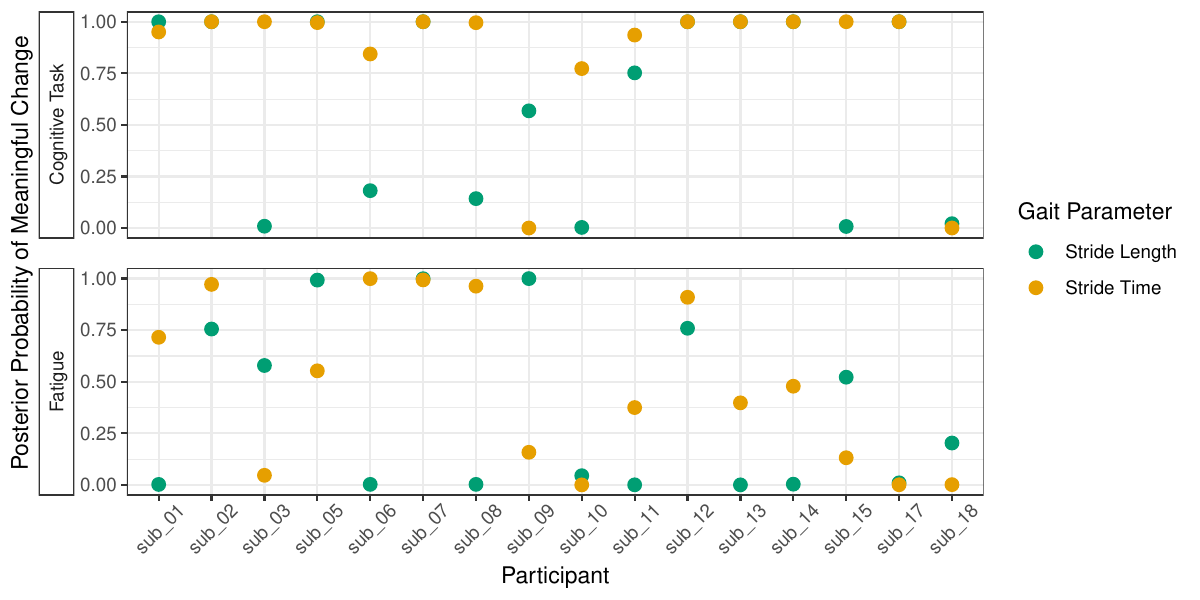}
    \caption{Posterior probabilities of meaningful effect for each participant and gait parameter. The probabilities of meaningful effect are heterogeneous among participants and between the two gait parameters.}
    \label{fig:posterior_prob_meaningful_change}
\end{figure}

To further quantify the marginal effects of fatigue and cognitive task on gait, the posterior probabilities of meaningful effects were estimated. The population-level analysis only showed small effects of fatigue on stride length and stride time. However, as illustrated in Figure~\ref{fig:posterior_prob_meaningful_change}, the posterior estimates for individual participants revealed highly heterogeneous gait effects among the participants, and the posterior probability of meaningful effect on the gait parameters were considerably high for some individuals (e.g., stride lengths for participants~5,~7 and~9, stride times for participants 2, 6, 7 and 8) compared to the others. Moreover, the pattern of probability for the two gait parameters also differed from person to person. The posterior probability of meaningful effect remained consistent between the two gait parameters for some participants~(e.g., participants~7,~10~and~17), while differed largely for some other participants~(e.g., participants~1,~6~and~8). Similar patterns could also be observed for the effects of cognitive task. 

\begin{figure}[h]
     \centering
     \includegraphics[width=0.9\textwidth]{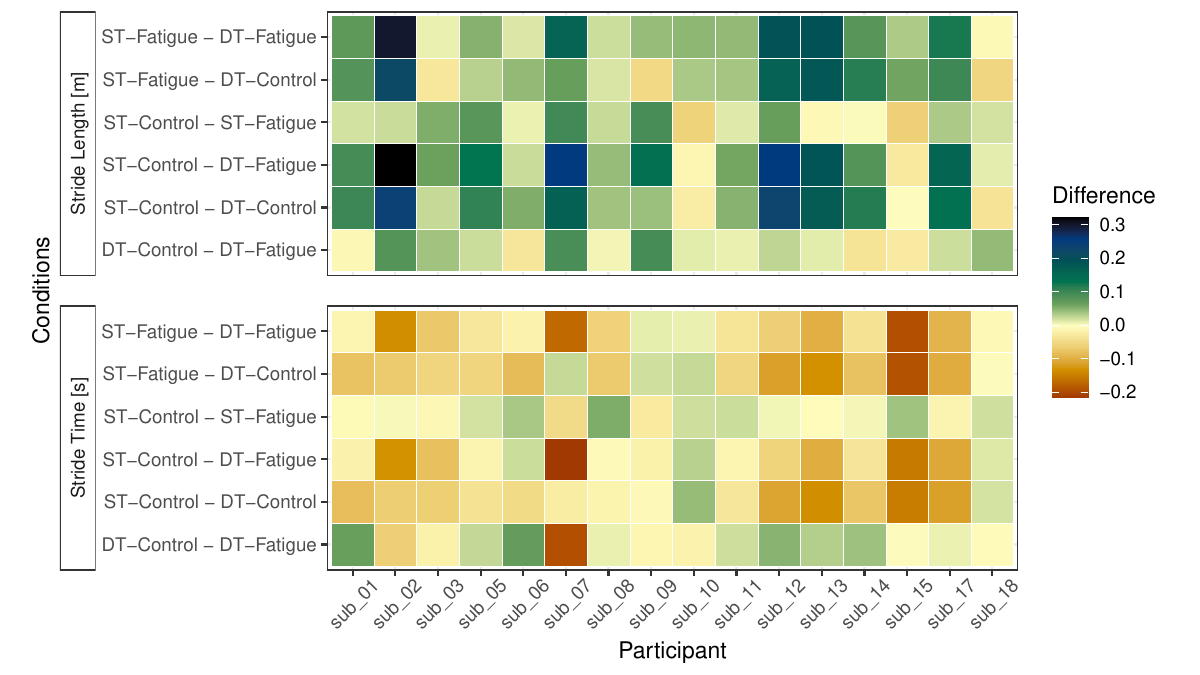}
    \caption{Heatmap of gait parameter differences between all combinations of condition pairs based on Posterior estimates of coefficients from the AR1 model. Each row represents the difference of the gait parameters between one pair of conditions. Gait changes between condition pairs are heterogeneous among participants.}
    \label{fig:heatmap_posterior_estimations}
\end{figure}

To provide a qualitative overview of the heterogeneous gait changes under the four walking conditions for each participant, we computed the difference between each pair of conditions using mean values under the four walking conditions obtained from the posterior distributions. As illustrated in Figure~\ref{fig:heatmap_posterior_estimations}, for the great majority of the condition pairs, the gait changes varied in both magnitude and direction among all participants.

In the sensitivity checks, we investigated how the results might change when different regression models (AR1, basic, time covariate) or different priors (informative and non-informative) were used. Overall, the AR1 and basic model had similar posterior distributions of parameters, and the time covariate model had a slightly shifted distribution. The difference between ST-Control and DT-Control for stride length, as represented by $\beta_2$, was similar in the AR1 model, the basic model and observed data, and larger compared to the posterior estimate from the time covariate model (-0.09 from observed data, the AR1 model and the basic model, -0.06 from time covariate model). The difference between ST-Control and ST-Fatigue, as represented by $\beta_3$, was negative in observed data and in the AR1 and basic model, but positive when estimated by the time covariate model (-0.03 from observed data, from the AR1 model and from the basic model, 0.03 from posterior estimate). Posterior estimates for stride times exhibited similar trends, in that the $\beta_2$ and $\beta_3$ estimates from the time covariate model were slightly different from those from the AR1 and the basic models.
A comparison of models with different priors indicated that they had no meaningful influence on the sampling results. A more detailed illustration of the effects of models and priors can be found in Supplementary Text C. Since no statistically significant differences in the posterior estimates using non-informative and informative priors were found, we focused on reporting and discussing results from models using non-informative priors.

\section{Discussions}
In this study, we re-analyzed data from a gait study, which was originally designed as a population-based trial, as single N-of-1 trials. 
Our observations from the posterior estimates demonstrate that heterogeneous responses do exist in our cohort. Especially for the effects of fatigue on gait, analyses based on N-of-1 trials are necessary to enable in-depth investigation of the individual effects that are not visible in the group-based ANOVA results. This is consistent with our prior knowledge that gait and gait responses to interventions are highly individual~\cite{HORST2017, chan2020effects, marxreiter2018sensor}, and the posteriors for each individual allow further investigation into causes underlying these individual effects. 

Our study introduced personalized analysis methods using example data from a healthy population. Nonetheless, spatiotemporal gait parameters such as stride length and stride time are commonly used to objectively and quantitatively measure gait performance in neurological diseases. In the example of Parkinson’s disease, gait impairment is often associated with a decrease in stride length and an increased risk of falling~\cite{bayle2016contribution}. Falls can result in injuries, fear of falling, and activity restriction, which are associated with increased institutionalization, reduced independence, and higher mortality rates. Previous studies have shown that by using various cueing strategies (e.g., visual or auditory cueing), the stride length could significantly increase up to that of control levels~\cite{schlick2016visual, pau2016effects}. A personalized approach could enable the identification of effective interventions for an individual to improve gait performance, and thus mitigate adverse consequences of gait impairment. Moreover, our method can also be generalized to investigate individual outcomes in a variety of settings.


When comparing group-based analyses with N-of-1 trials approaches, it is worth noting that the covariate effects are treated differently. Typical group-level models statistically control for the effects of covariates related to the traits of participants, such as age, height and weight. In contrast, the N-of-1 approach automatically adjusts for all baseline covariates, since each person always has the same baseline covariate values.


 The dataset used in our study was unconventional for N-of-1 trials methods in the sense that it was obtained from a study originally designed for population-level analyses. In contrast to the multiple crossovers for typical N-of-1 trials, each participant was measured only in one session for the baseline, underwent the intervention once, and the intervention effects were measured in a subsequent session. Hundreds of data points (gait cycles) were measured under both conditions, which are sufficient for statistical analysis on a single person. Nevertheless, in order to enable the analysis of the data through the lens of N-of-1 trials, several assumptions were made with respect to time-dependent effects, carryover effects, and effects of task/treatment order. 

In the following, we discuss these assumptions and the ramifications of not having additional crossovers in our study design.
Regarding time-dependent effects, we assumed that the one-week break between the ST and DT visits did not induce any effect on the gait characteristics of an individual. Only in this case, the effect observed from DT condition could be attributed to cognitive task and not with time as a confounder. We based our assumption on evidence that an individual's gait characteristics are persistent over a long period of time~\cite{HORST2017}. For other types of outcomes that fluctuate over time or are more sensitive to uncontrolled factors, the effect of time between visits should not be neglected.

The order of ST and DT visits was randomized among all participants. However, with only one crossover, the order was fixed for the same person. In addition, the order of the control and fatigue conditions during both visits was fixed by the experimental design.
This lack of within-person randomization may become problematic when two types of carryover effects exist: 1) carryover from the first visit to the second visit, and 2) carryover within one visit, from the first walking session to the second walking session. Regarding the first scenario, the one-week break between two visits can be considered a washout period, where the effects of fatigue exercise from the previous visit are sufficiently removed. Before the second visit, we confirmed with each participant that they did not feel any effects of the fatigue exercise. Regarding the second scenario, it is only reasonable to keep the order of first non-fatigued walk, then fatigued walk during the same visit, and there were no obvious carryover effects from the non-fatigued state to the fatigued state. When generalizing our methods to other data from population-based trials, it is important to examine all possible carryover effects, and evaluate to which extent these effects will affect the analysis outcome.

In the case of prominent carryover effects, the single-crossover design does not isolate the carryover effects from the intervention effects for the individual. As one approach, the carryover effects could be modeled in the analysis to still allow efficient and unbiased estimation of the effects~\cite{gartner2022comparison}. In future work, additional crossovers between the fatigue and dual-task conditions can be added to the study design, in order to introduce randomization within one person. 

In addition to the time-dependent effect and carryover effect, the task/treatment order may have an influence on motivation and habituation to the task. Furthermore, events that alter a participant’s gait, such as injury, may occur between visits. Although no evidence of such an effect was observed in our example study, future studies should consider these factors when analyzing trials with fixed treatment orders. 

In our analyses, informative and non-informative priors did not have an effect on posteriors for the same model, as illustrated in Supplementary~Figure~13. One possible reason could be that the values of the non-informative priors were similar to those of the informative priors. More concretely, only prior knowledge of the mean and standard deviation for the baseline (ST-Control) was introduced, and these values (centered around 1 with very small standard deviations) were close to the non-informative priors (centered around 0 with a standard deviation of $10^{-3}$). It is therefore worth emphasizing that for use cases where the informative priors differ largely from non-informative default priors, an informed choice of priors might affect the posterior estimates to a larger degree and cannot be ignored. Visualizations similar to our Supplementary~Figure~13 are helpful for qualitative comparison between different priors.

Posterior estimates of the model parameters from the AR1 and basic models matched the distributions of the observed values, whereas slight deviations could be observed for posterior estimates of the time covariate model. Moreover, for four participants (\#2, 9, 12, 18), the MCMC chains did not converge for the model parameter $\beta_5$ which was associated with the linear representation of time in the design matrix. In our opinion, these observations indicate that the assumption of a linear time effect with the time covariate model does not accurately represent the data. As discussed by Heckenstenden~et~al.~\cite{hecksteden2015individual}, repeated measurements during a single uninterrupted intervention period could be used as a surrogate for repeated interventions, however, it is reasonable to assume autocorrelation between measurements, and non-linear adaptation may occur during the measurement period. Our study indicates that in such settings, the effects of time are more appropriately modeled with the temporal autocorrelation described by the AR1 model. With the assumption of stationarity, our AR1 model describes a simple pattern of correlation that declines linearly according to the time lag, but is independent of the actual time of the observation. In this regard, samples from all four walking conditions were concatenated and modeled with the same correlation structure, which yielded a good model fit. Nonetheless, it is worth noticing that the MCMC chains failed to converge for the AR1 models for two participants (\#7, 13) for stride length. During data collection, we observed that the general gait patterns of these two participants were particularly affected by the interventions, and initial data exploration revealed a large variability in gait parameters. We assume that the true data distributions of their gait parameters are different from those of the other participants, and the true relationship between the variables is non-linear. In this case, a more flexible model with a non-linear structure could be better suited for analysis.

As future work, the effect sizes of interventions can be estimated and further investigated based on the posteriors obtained from the Bayesian analyses~\cite{kelter2020analysis}. Aggregated N-of-1 trials analyses can be performed to investigate the underlying causes of the personalized responses to intervention ~\cite{gartner2022comparison}. In our study, the different responses could potentially be associated with the participants' pre-existing health conditions, anthropometric features or stable lifestyle habits (e.g., as measured by the IPAQ questionnaire), or a combination of all these factors. In future studies, larger and more heterogeneous cohorts will provide additional features for analysis. Based on these findings, personalized advice or interventions could reduce the risk of falls or injury for vulnerable individuals. 

\section{Conclusion}
Our study provides an example of how to initiate an in-depth investigation of treatment effects on an individual level using data from population-level studies. We demonstrate the use of Bayesian models to study individual-level effects of interventions, and point out aspects to consider for future studies.

\noindent {\bf{Conflict of Interest}}\\
\noindent {The authors have declared no conflict of interest.}

\noindent {\bf{Acknowledgement}}\\
\noindent The authors would like to thank all participants in this study. We would also like to thank Urs Granacher and Clemens Markus Brahms for their support in developing the study. 
This study received funding from the Deutsche Forschungsgemeinschaft (DFG, German Research Foundation) – project number 491466077, and has been partly funded by the Federal Ministry of Education and Research of Germany in the framework of KI-LAB-ITSE (project number 01IS19066).

\bibliographystyle{unsrt}

\bibliography{references}

\newpage
\section*{Supplementary materials}
\appendix
\section{Details on research methods}
\label{suppl_text_1}
\subsection{Fatigue protocol and cognitive task}
\label{suppl_cognitive_task}
During the fatigue protocol, the participants wore a weighted vest matched to 30 \% of their body weight, and repeatedly stood up from a chair at a self-selected, rapid pace until failure. Fatigue was assessed using the Borg Rating of Perceived Exertion (referred to as the Borg scale in the following text)~\cite{borg2006comparison}, as well as blood lactate concentration measured using blood sampled from the earlobe~\cite{finsterer2012biomarkers}. The threshold for fatigue was Borg scale of 15, and all of the participants reported higher than this threshold. We did not set a threshold value for blood lactate under the fatigue condition, because the underlying physiological factors for different people may lead to very different measurements. However, we confirmed that all the participants had a blood lactate concentration baseline (before the exercise) below 2 mmol/L, which is in agreement with previously reported levels at rest~\cite{miller2002lactate}. In addition, the t-test indicated a statistically significant increase in blood lactate concentration among all participants following the fatigue protocol. The p-values were less than $10^{-5}$ for both single- and dual-task conditions.

The cognitive task required participants to serially subtract the number seven from a 4-digit number between 3000 and 9000 provided by the experimenter, which was randomly generated anew for each participant. Participants were asked to verbalize the numbers (e.g., 3745, 3738, 3731, [...]), so that the answers could be documented with the audio recorder and analyzed later. In order to minimize the learning effect, where participants become increasingly familiar with the number counting task during walking and subsequently alter their gait pattern within the session, we instructed them to first engage in a~6-minute practice dual-task walk. This allowed for the learning effect to primarily occur during the practice phase, resulting in more consistent gait patterns during the actual recording.

The study was approved by the ethics committee of the University of Potsdam (63/2020), and all experiments were conducted according to the latest revision of the Declaration of Helsinki. All participants provided written consent prior to data collection.

\subsection{IMU gait analysis}
\label{appendix_gait_analysis}
Two IMUs (Physilog\textsuperscript{\textregistered}5, Gait~Up, Switzerland) were attached to the left and right insteps of the participants. Tri-axial acceleration and angular velocity data were recorded at 128 Hz, and the acceleration- and gyro ranges were $\pm~16$~g and $\pm~1000$~dps, respectively. An audio recorder was attached close to the left collarbone in order to document the responses of the cognitive tasks recorded under the dual task condition.

Stride length and stride time of all strides from the walking sessions were extracted from the raw IMU data using an error-state Kalman filter-based algorithm, which utilizes zero-velocity update (ZUPT) to correct for accumulating drift errors caused by inertial integration. The algorithm has been described in detail, and validated against gold standard reference systems in previous studies~\cite{tunca2017inertial, zhou2020}. Briefly, tri-axial acceleration and angular velocity signals were used as inputs, the stance phases were identified with gyro magnitude threshold, and an error-state Kalman filter was employed to periodically correct the drift during stance phases, thus enabling estimation of the three-dimensional trajectories of the feet movement. The gait events (toe-off, initial contact) were identified using signal features in the gyroscope data. Temporal gait parameters, such as swing time and stance time, were derived directly from the gait events. Spatial gait parameters, such as stride length and clearance, were obtained by segmenting the trajectories into individual strides. Turning strides at the ends of the walkway were excluded based on a threshold for the change in foot orientation. Acceleration and deceleration phases were excluded by removing two strides before and after the actual turning strides. Additional outlier strides were defined as those whose gait parameter values were larger or smaller than three standard deviations around the mean (effectively 0.3\% of the data) and were excluded from further analyses.

The above-described methods extract stride length and stride time from IMU data. Alternatively, it is also possible to measure these gait parameters using other typical clinical gait analysis setups such as walkways with pressure sensors or multi-camera systems.

\subsection{Data distribution after down-sampling}

\begin{figure}
    \centering
    \includegraphics[width=0.9\textwidth]{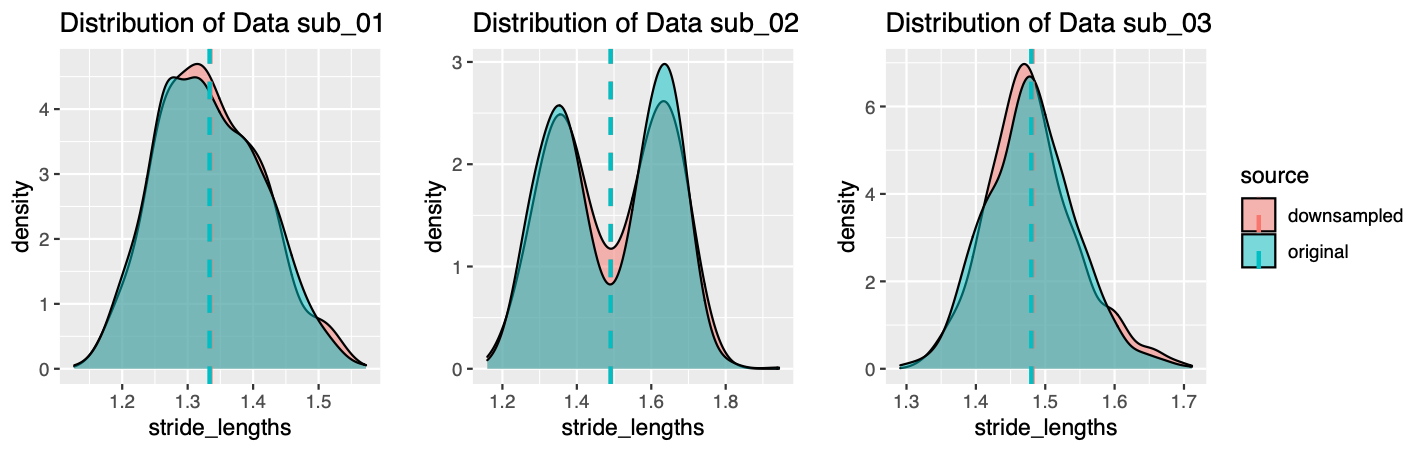}
    \caption{Density plots of data points from three example participants. Dashed lines indicate mean values. Down-sampling by a factor of 5 did not change the overall distribution of the data.}
    \label{fig:SL_dist_downsample}
\end{figure}

\begin{figure}
    \centering
    \includegraphics[width=0.6\textwidth]{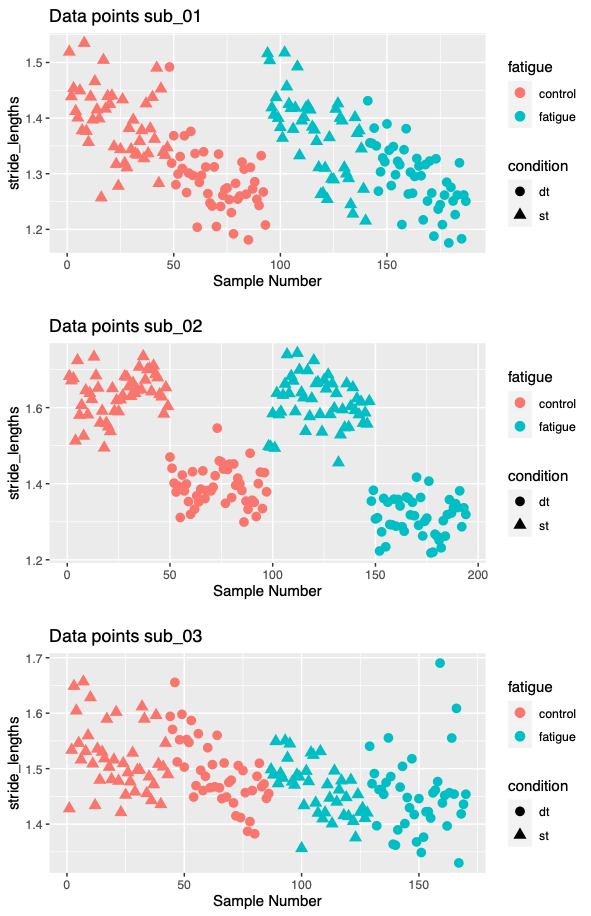}
    \caption{Scatter plots of data points from three example participants. There is no obvious change in the gait parameters over time with each walking condition for the same person.}
    \label{fig:SL_over_time}
\end{figure}

We confirmed with visual inspection that down-sampling by a factor of 5 did not change the overall distribution of the data, as shown in Figure~\ref{fig:SL_dist_downsample}. In addition, there is no obvious change in the gait parameters over time with each walking condition for the same person, as shown in Figure~\ref{fig:SL_over_time}. 

\newpage
\section{Details on statistical models}
\label{suppl_text_2}
\subsection{Population-based model}
The population-based ANOVA model is defined as follows:

\begin{equation}
    \label{eq:anova}
    y_{ijk} = \mu + \alpha_i + \beta_j + \gamma_{ij} + \delta_1x_{1k} + \delta_2x_{2k} + \delta_3x_{3k} + \epsilon_{ijk}
\end{equation}

\noindent where $y_{ijk}$ is the dependent variable, which is the observed gait parameter value for the $i$th level of within-subject factor "dual-task" and the $j$th level of within-subject factor "fatigue" for the $k$th subject. Note that each of the two factors $i$ and $j$ has two levels, namely, with or without dual-task, and with or without fatigue, respectively.
 $\mu$ is the grand mean of all observations.
 $\alpha_i$ is the main effect of factor "dual-task", which represents the difference between the mean of the $i$th level of factor "dual-task" and the grand mean.
 $\beta_j$ is the main effect of factor "fatigue", which represents the difference between the mean of the $j$th level of factor "fatigue" and the grand mean.
 $\gamma_{ij}$ is the interaction effect between factors "dual-task" and "fatigue".
 $\delta_1$, $\delta_2$, and $\delta_3$ are the coefficients for between-subjects covariates "age", "body height", and "body mass", respectively. These terms represent how changes in these covariates are associated with changes in the dependent variable.
 $x_{1k}$, $x_{2k}$, and $x_{3k}$ are the values for between-subject covariates "age", "body height", and "body weight", respectively, for the $k$th subject.
 $\epsilon_{ijk}$ is the residual error term, which represents random variation not accounted for by any other terms in the model.

For calculating the two-way repeated measures ANOVA we used the function \texttt{ezANOVA} \cite{ezanova} from the \texttt{ez} package, and reported the generalized $\eta^2$ (\texttt{ANOVA.ges}) \cite{Olejnik2003} output as the effect size, which is recommended for repeated measures designs\cite{Bakeman2005}. Generalized $\eta^2$ can be interpreted analogously to Cohen's recommendations \cite{Cohen2013, Lakens2013}.\\

\subsection{Bayesian models}
We used Bayesian repeated-measures models to fit probabilistic models of the data distribution for the four walking conditions. The models provide a probabilistic description of the data for interpretation~\cite{makowski2019indices}. The Markov Chain Monte Carlo (MCMC) method with Gibbs sampling was used to estimate parameters describing the data distribution. 
MCMC performs Monte Carlo integration by drawing samples from a probability distribution with the construction of a Markov chain, where the value of the current variable is dependent on the value of the previous variable in the chain. With sufficient numbers of steps, the sampled values will converge to the value that is being inferred~\cite{geyer1992practical}. Gibbs sampling is one of the most commonly used MCMC algorithms. It draws samples for each parameter from the full conditional distributions of that parameter~\cite{smith1993bayesian}.

The basic model with non-informative priors is specified as follows. (More details can be found, for example, in a tutorial on factorial ANOVA implementation using JAGS\footnote{https://agabrioblog.onrender.com/tutorial/factorial-anova-jags/factorial-anova-jags/, last retrieved on 2023-07-20}.) \\
\begin{scriptsize}
\begin{verbatim}
modelString = " 
    model {
      #Likelihood
      for (i in 1:n) {
      y[i]~dnorm(mean[i],tau)
      mean[i] <- inprod(beta[],X[i,])
      }
      #Priors
      beta[1] ~ dnorm(0,1.0E-3)
      for(i in 2:ngroups) {
      beta[i] ~ dnorm(0,1.0E-3)
      }
      sigma ~ dunif(0, 100)
      tau <- 1 / (sigma * sigma)
    }
"
\end{verbatim}
\end{scriptsize}

The primary outcome of this study was the distribution of gait parameters under the four walking conditions, which can be derived from the model parameters. The JAGS program used in this work uses a dialect of the BUGS modeling language. In BUGS language, the normal distribution is parameterized in terms of precision (\textit{tau}), which is the inverse of variance (\textit{sigma} squared)~\cite{plummer2017jags}.
In the model string, the parameter \textit{n} in the likelihood model represents the total number of data points used for the simulation, whereas the \textit{ngroups} in priors represents a total number of elements in the beta vector, namely, the number of columns in the design matrix \textit{X}.

The time covariate model was the same as the basic model, except that the design matrix \textit{X} had an additional column with incremental integers for each walking condition, which represents the time component.
The AR1 covariance model with non-informative priors was defined as follows. The definition of the half-Cauchy distribution was adopted from Gelman~\cite{gelman2006prior}. The posterior probability of meaningful effects on gait parameters was estimated using the same AR1 model with non-informative priors, the percentage threshold was 3\% for stride length, and 2\% for stride time. The method for estimating the probability was adapted from Stunnenberg~et~al.~\cite{stunnenberg2018effect}. The derivations of the marginal effects of fatigue and cognitive task are detailed in the following section. \\

\begin{scriptsize}
\begin{verbatim}
modelString = "
    model {
        #Likelihood
        for (i in 1:n) {
        mean[i] <- inprod(beta[],X[i,])
        }
        y[1:n] ~ dmnorm(mean[1:n],Omega)
        for (i in 1:n) {
        for (j in 1:n) {
        Sigma[i,j] <- sigma2*(1- phi*phi)*(equals(i,j) + (1-equals(i,j))*pow(phi,abs(i-j))) 
        }
        }
        Omega <- inverse(Sigma)
        
        #Priors
        phi ~ dunif(-1,1)
        beta[1] ~ dnorm(0,1.0E-3)
        for(i in 2:ngroups) {
        beta[i] ~ dnorm(0,1.0E-3)
        }
        sigma <- z/sqrt(chSq)    # prior for sigma; cauchy = normal/sqrt(chi^2)
        z ~ dnorm(0, 0.16)I(0,)  # positive part of normal distribution, Cauchy scale = 2.5
        chSq ~ dgamma(0.5, 0.5)  # chi^2 with 1 d.f.
        sigma2 = pow(sigma,2)

        # Probability effects of fatigue and cognitive task
        threshold <- 0.03    # threshold for meaningful effect
        # is effect ((6) in B.3) larger than 3% of non-fatigued gait (5(b) in B.3):
        diff_fatigue <- abs(beta[3] + 0.5*beta[4]) - (beta[1] + 0.5*beta[2])*threshold 
        diff_cognitive_task <- abs(beta[2] + 0.5*beta[4]) - (beta[1] + 0.5*beta[3])*threshold
        p_fatigue <- step(diff_fatigue)    # probability of larger than the threshold
        p_cognitive_task <- step(diff_cognitive_task)
    }
"
\end{verbatim}
\end{scriptsize}

\subsection{Marginal effects of fatigue and cognitive task}
Along with the AR1 model parameters, we estimated the posterior probability of meaningful effects of fatigue on gait parameters. In this section, we demonstrate how the marginal effects of fatigue and cognitive task are estimated from the $\bm{\beta}$ coefficients from the AR1 model.

\begin{enumerate}
    \item Let $W = 1$ when a participant has been exposed to fatigue (i.e., fatigue condition), and $W = 0$ otherwise (i.e., control condition).
    \item Let $V = 1$ when a participant has performed the cognitive task (i.e., dual-task condition), and $V = 0$ otherwise (i.e., single-task condition).
    \item To estimate the marginal effects, we will use $E(Y|W) = \sum_v E(Y|W,V=v)Pr(V=v)$ and similarly for $E(Y|V)$. This follows from: first, $E(Y|W) = E\{E(Y|W,V)|W\}$ by the law of total expectation (i.e., generalized Adam's law). Further, $Pr(V|W = w) = Pr(V)$ for $w \in \{0, 1\}$ due to our study design in which the probability of performing the cognitive task is fixed and independent of the assignment of the fatigue condition. With this, $E(Y|W) = E\{E(Y|W,V)|W\} = E\{E(Y|W, V)\} = \sum_v E(Y|W,V=v)Pr(V=v)$.
    \item For the model in Equation (1), we have:
    \begin{enumerate}
        \item $E(Y|W=0,V=0) = \beta_1$
        \item $E(Y|W=0,V=1) = \beta_1 +\beta_2$
        \item $E(Y|W=1,V=0) = \beta_1 +\beta_3$
        \item $E(Y|W=1,V=1) = \beta_1 +\beta_2 +\beta_3 +\beta_4$
    \end{enumerate}
    \item Hence, we have:
    \begin{enumerate}
        \item Mean gait when exposed to fatigue: \\
        $E(Y|W = 1) = E(Y|W = 1, V = 0) Pr(V = 0) + E(Y|W = 1, V = 1) Pr(V = 1)$ \\
        $= (\beta_1 +\beta_3)Pr(V=0) + (\beta_1 + \beta_2 + \beta_3 + \beta_4)Pr(V=1)$ \\
        $= (\beta_1 +\beta_3 )\{Pr(V=0)+Pr(V=1)\} +(\beta_2 +\beta_4)Pr(V=1)$ \\
        $= (\beta_1 +\beta_3) +(\beta_2 +\beta_4)Pr(V=1)$ \\

        \item Mean gait when not exposed to fatigue: \\
        $E(Y|W = 0) = E(Y|W = 0, V = 0) Pr(V = 0) + E(Y|W = 0, V = 1) Pr(V = 1)$ \\
        $= \beta_1Pr(V=0) + (\beta_1 +\beta_2)Pr(V=1)$ \\
        $= \beta_1\{Pr(V=0)+Pr(V=1)\}+\beta_2Pr(V=1)$ \\
        $= \beta_1 +\beta_2 Pr(V=1)$ \\
    \end{enumerate}
    \item Here, $Pr(V=0)$ = $Pr(V=1)$ = 0.5 because the participant performed the cognitive task half of the time. Hence, the marginal effect of fatigue on gait is: \\
    $E(Y|W = 1) - E(Y|W = 0)$ \\
    $= \{(\beta_1 +\beta_3)+(\beta_2 +\beta_4)Pr(V=1)\} - \{\beta_1 +\beta_2 Pr(V=1)\}$ \\
    $= (\beta_1 +\beta_3) - \beta_1 +(\beta_2 + \beta_4 - \beta_2)Pr(V=1)$ \\
    $= \beta_3 + \beta_4 Pr(V=1)$ \\
    $= \beta_3 + 0.5\beta_4$ \\
\end{enumerate}


Similarly, to obtain the marginal effect of cognitive task, we use the same conditional probabilities under the four walking conditions derived in the previous steps,
\begin{enumerate}
    \item Hence, we have:
    \begin{enumerate}
        \item Mean gait when exposed to cognitive task: \\
        $E(Y|V = 1) = E(Y|V = 1, W = 0) Pr(W = 0) + E(Y|V = 1, W = 1) Pr(W = 1)$ \\
        $= (\beta_1 +\beta_2)Pr(W=0) + (\beta_1 + \beta_2 + \beta_3 + \beta_4)Pr(W=1)$ \\
        $= (\beta_1 +\beta_2 )\{Pr(W=0)+Pr(W=1)\} +(\beta_3 +\beta_4)Pr(W=1)$ \\
        $= (\beta_1 +\beta_2) +(\beta_3 +\beta_4)Pr(W=1)$ \\

        \item Mean gait when not exposed to cognitive task: \\
        $E(Y|V = 0) = E(Y|V = 0, W = 0) Pr(W = 0) + E(Y|V = 0, W = 1) Pr(W = 1)$ \\
        $= \beta_1Pr(W=0) + (\beta_1 + \beta_3)Pr(W=1)$ \\
        $= \beta_1\{Pr(W=0)+Pr(W=1)\} + \beta_3Pr(W=1)$ \\
        $= \beta_1 +\beta_3 Pr(W=1)$ \\
    \end{enumerate}
    \item Here, $Pr(W=0)$ = $Pr(W=1)$ = 0.5 because the participant performed the fatigue exercise half of the time. Hence, the marginal effect of cognitive task on gait is: \\
    $E(Y|V = 1) - E(Y|V = 0)$ \\
    $= \{(\beta_1 +\beta_2)+(\beta_3 +\beta_4)Pr(W=1)\} - \{\beta_1 +\beta_3 Pr(W=1)\}$ \\
    $= (\beta_1 + \beta_2)+\{(\beta_3 +\beta_4)-(\beta_1 +\beta_3)\}Pr(W=1)$ \\
    $= (\beta_1 + \beta_2) - \beta_1 +\{(\beta_3 +\beta_4)-\beta_3\}Pr(W=1)$ \\
    $= \beta_2 + \beta_4 Pr(W=1)$ \\
    $= \beta_2 + 0.5\beta_4$ \\
\end{enumerate} 

\newpage
\section{Details on quality control}
\label{suppl_text_3}
\subsection{MCMC chain convergence}
Convergence of the MCMC chain was confirmed with visualization using trace plots and the convergence statistic potential scale reduction factor (PSRF). The trace plot displays sampled values over a number of iterations for each chain and each model parameter. Stable and uniform patterns (i.e., a horizontal band with no particular patterns) for both chains indicate convergence. The PSRF is an estimated factor by which the current distribution of the parameter might be reduced if the simulations were to continue for an infinite number of iterations~\cite{gelman1992inference}. The PSRF plot shows the median and upper confidence limits (confidence~=~0.95) against the number of iterations. An upper limit close to 1 indicates approximate convergence, as the current distribution is no longer over-dispersed in respect to the target distribution. In our study, both the trace plots and PSRF confirmed chain convergence for all simulations for stride length. Supplementary Figures~7~and~8 show example trace plots and PSRF plots of converged chains, respectively. For stride time, the chains from the AR1 models did not converge for participants sub\_07 and sub\_13 for all model parameters. Chains from the time covariate models did not converge for participants sub\_02, sub\_09, sub\_12, and sub\_18 for the parameter $\beta_5$, which was associated with the time component in the design matrix. Examples of trace plots and PSRF for non-convergence can be found in Supplementary Figures~9~and~10.

\begin{figure}[h]
     \centering
     \includegraphics[width=0.9\textwidth]{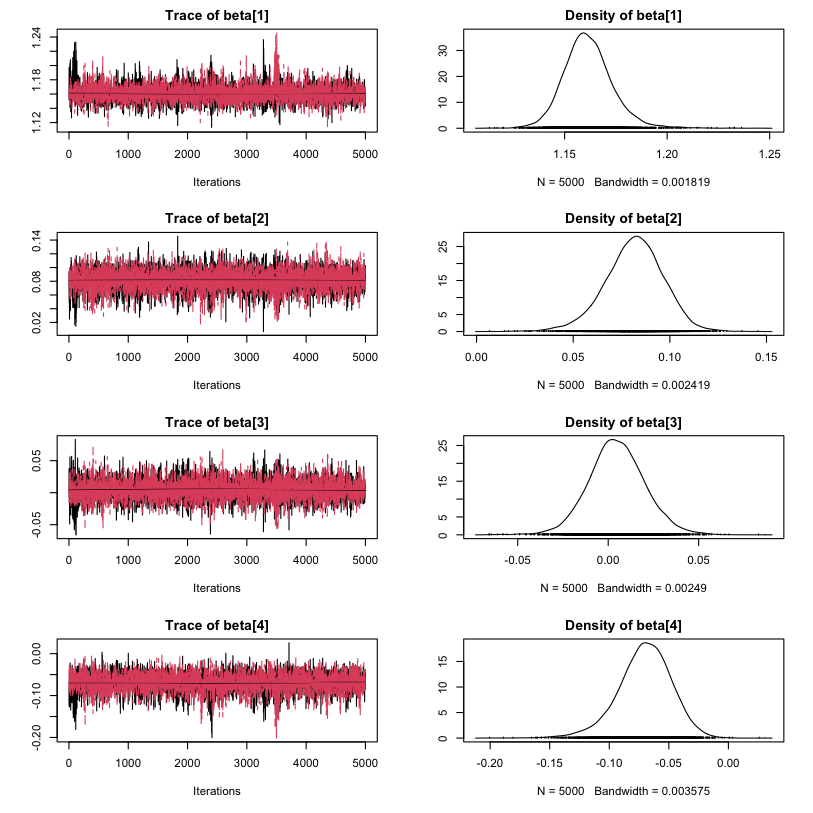}
     \caption{Example trace plots and density plots for MCMC chain convergence diagnosis for the AR1 model with data from sub\_01 stride length. The plots of chains that have converged are similar for other models and their parameters described in this study.}
     \label{fig:traceplot_SL_sub_01}
 \end{figure}
 
 \begin{figure}[h]
     \centering
     \includegraphics[width=0.9\textwidth]{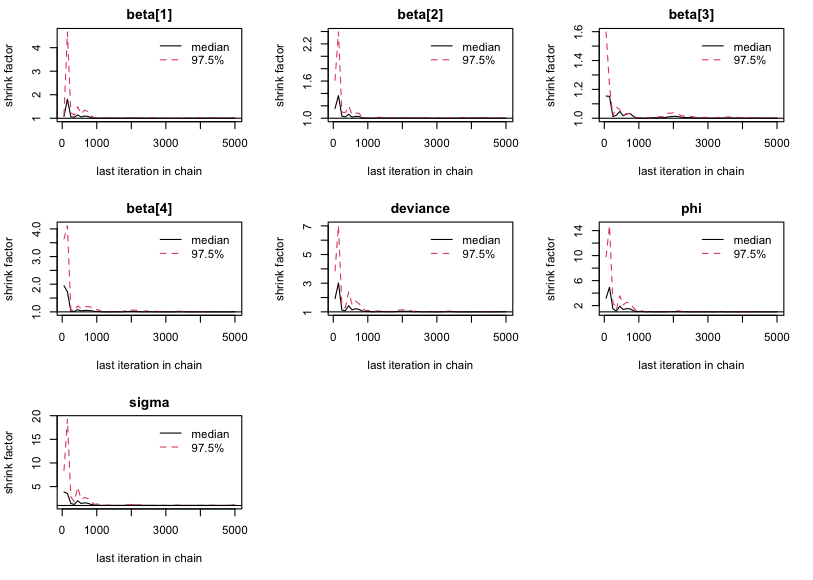}
     \caption{Example Potential scaling reduction factor (PSRF) plots for MCMC chain convergence diagnosis for the AR1 model with data from sub\_01 stride length. The plots of chains that have converged are similar for other models and their parameters described in this study.}
     \label{fig:psrf_plot_SL_sub_01}
\end{figure}
 
 \begin{figure}[h]
     \centering
     \includegraphics[width=0.9\textwidth]{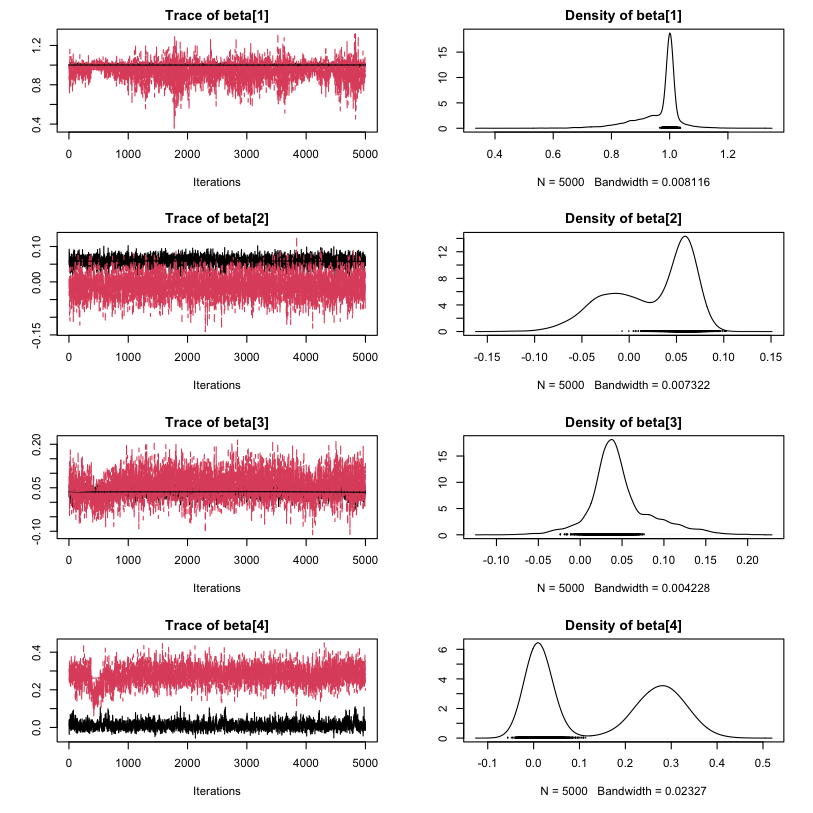}
     \caption{Example trace plots and density plots for MCMC chain convergence diagnosis for the AR1 model with data from sub\_07 stride length. The plots of chains that are not converged are similar for other models and their parameters described in this study.}
     \label{fig:traceplot_ST_sub_07}
 \end{figure}

\begin{figure}[h]
     \centering
     \includegraphics[width=0.9\textwidth]{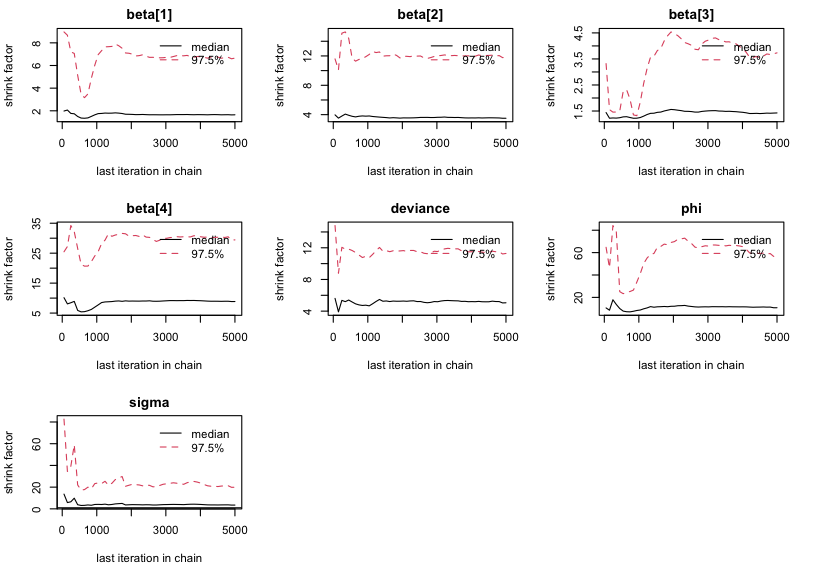}
     \caption{Example Potential scaling reduction factor (PSRF) plots for MCMC chain convergence diagnosis for the AR1 model with data from sub\_07 stride length. The plots of chains that are not converged are similar for other models and their parameters described in this study.}
     \label{fig:psrf_plot_ST_sub_07}
\end{figure}

\subsection{MCMC chain resolution}
The resolution of the MCMC chain was measured with effective sample size (ESS). A higher ESS indicates more information content, or higher effectiveness of the sample chain. In cases where observed data samples are highly autocorrelated, the ESS might be relatively small compared to the total sample size. Supplementary file for posterior estimates at \url{https://github.com/HIAlab/gait_nof1trials/wiki/Data} shows the ESS for each model and parameter (n.eff). 

\subsection{Posterior predictive check}

\begin{figure}[h]
    \centering
    \includegraphics[width=\textwidth]{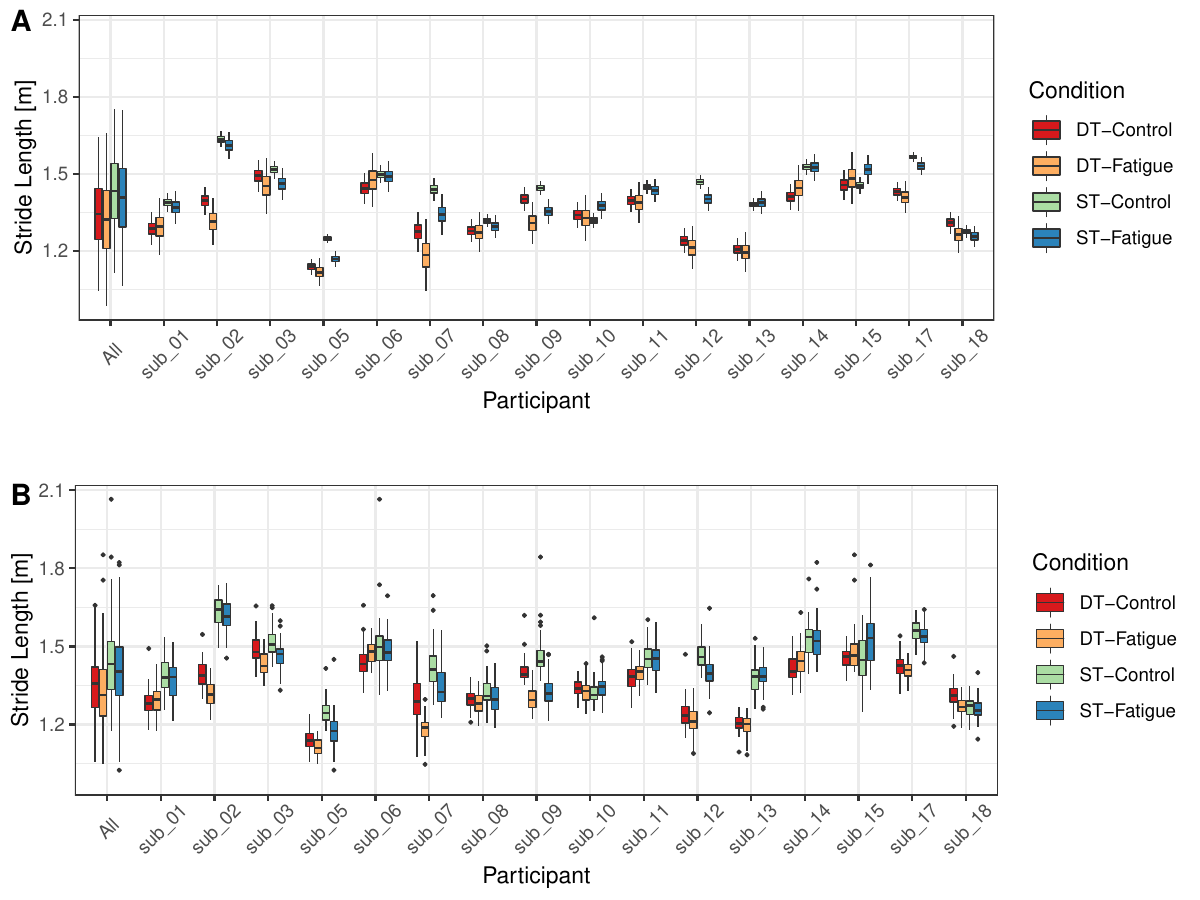}
    \caption{Comparison of (A) posterior estimates and (B) observed values for stride length.}
    \label{fig:boxplot_post_predictive_check_SL}
\end{figure}

\begin{figure}[h]
    \centering
    \includegraphics[width=\textwidth]{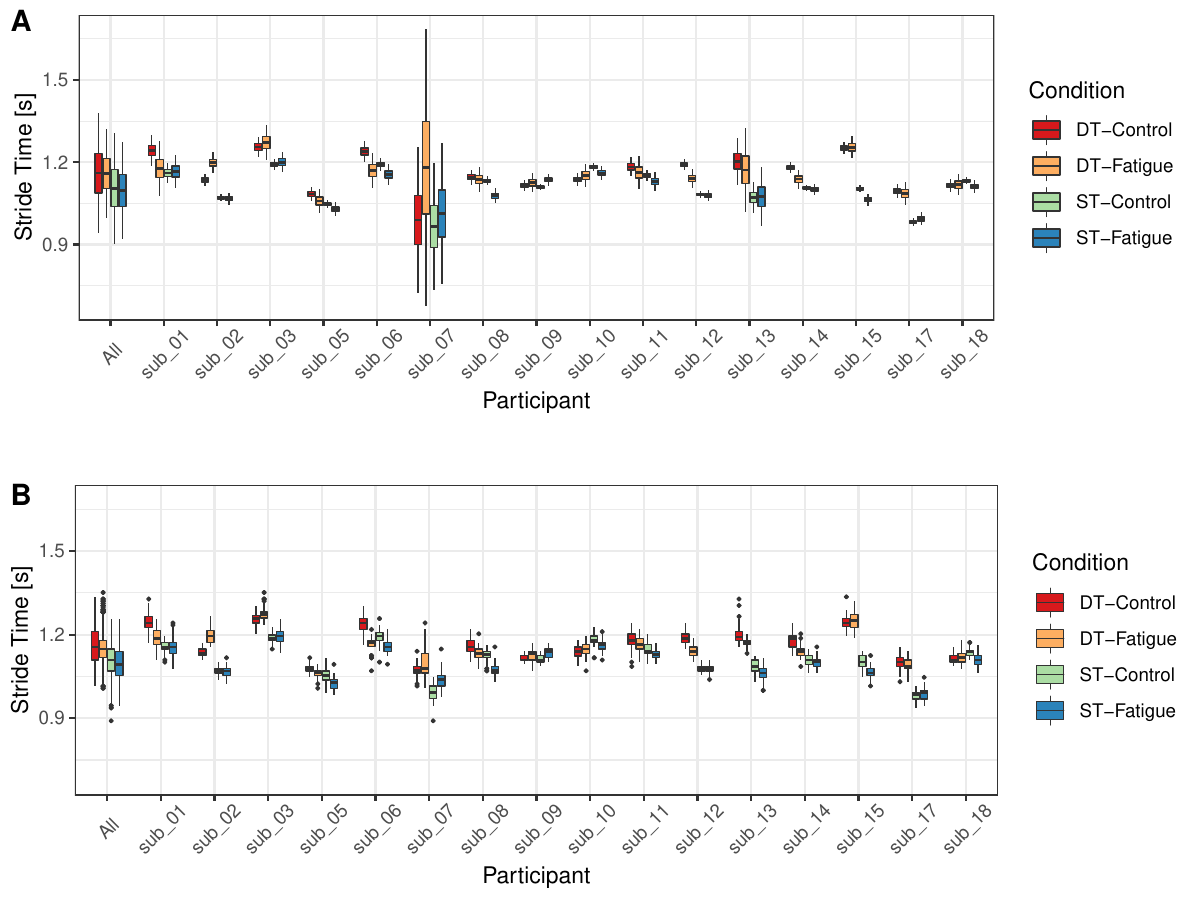}
    \caption{Comparison of (A) posterior estimates and (B) observed values for stride time.}
    \label{fig:boxplot_post_predictive_check_ST}
\end{figure}

Figure~\ref{fig:boxplot_post_predictive_check_SL} and Figure~\ref{fig:boxplot_post_predictive_check_ST} show that for both stride length and stride time, posterior estimation of the data distributions for each participant and each experimental condition closely resembles the observed data distributions. 

\subsection{Compare models}
\begin{figure}[h]
    \centering
    \includegraphics[width=0.9\textwidth]{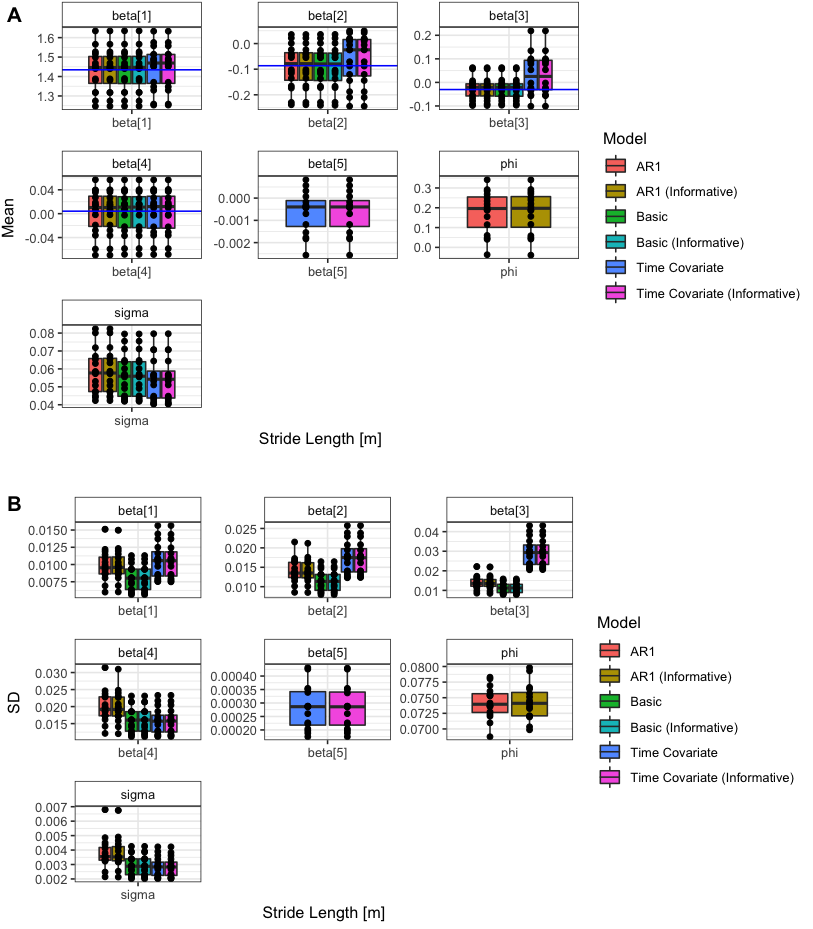}
    \caption{Compare (A) mean and (B) standard deviation of posterior distributions from different models for stride length. Each data point in the boxplot represents the aggregated value from one participant. Blue horizontal lines indicate the mean values of the parameter calculated from the observed data, not all Bayesian model parameters could be directly calculated from the observed data. No effect of informative priors could be observed, whereas different models produced slightly different posteriors. Only posteriors whose chains have converged are plotted here.}
    \label{fig:compare_models_posterior_SL}
\end{figure}

\begin{figure}[h]
    \centering
    \includegraphics[width=0.9\textwidth]{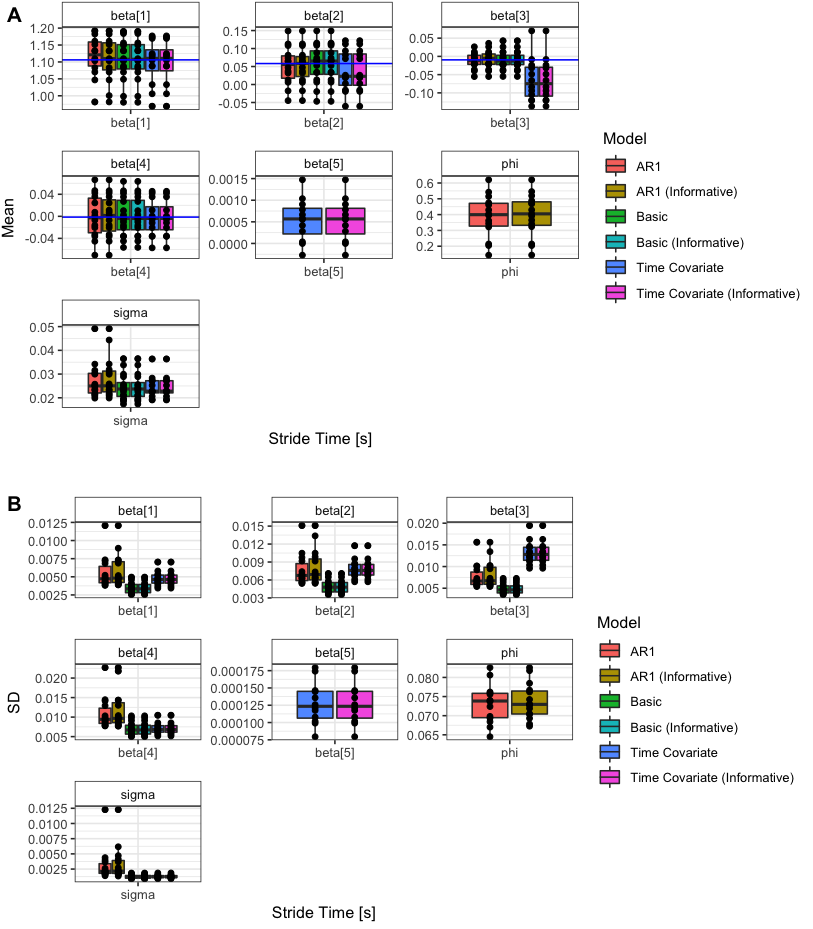}
    \caption{Compare (A) mean and (B) standard deviation of posterior distributions from different models for stride time. Each data point in the boxplot represents the aggregated value from one participant. Blue horizontal lines indicate the mean values of the parameter calculated from the observed data, not all Bayesian model parameters could be directly calculated from the observed data. No effect of informative priors could be observed, whereas different models produced slightly different posteriors. Only posteriors whose chains have converged are plotted here.}
    \label{fig:compare_models_posterior_ST}
\end{figure}

Posteriors of the main model (AR1) and two alternative models (basic and time covariate) were plotted in combination with their corresponding non-informative and informative priors for comparison. Figure~\ref{fig:compare_models_posterior_SL} and Figure~\ref{fig:compare_models_posterior_ST} show means and standard deviations for posteriors of stride length and stride time, respectively. Informative priors did not have a visible influence on posteriors, whereas the three different models produced slightly different posteriors.

\end{document}